\documentclass[twocolumn]{aastex6}
\usepackage{graphicx}
\usepackage{natbib}
\usepackage{amsmath,amsthm,amssymb}
\usepackage{color}
\usepackage{ulem}
\usepackage{epstopdf}
\newcommand{\ynirc}{\ensuremath{Y_{\rm NIRC2}}}
\newcommand{\jmko}{\ensuremath{J_{\rm MKO}}}
\newcommand{\hmko}{\ensuremath{H_{\rm MKO}}}
\newcommand{\kmko}{\ensuremath{K_{\rm MKO}}}
\newcommand{\mua}{\ensuremath{\mu_\alpha{\rm cos}\,\delta}}     
\newcommand{\mud}{\ensuremath{\mu_\delta}}                                

\newcommand{\my}{\protect \hbox {mas yr$^{-1}$}}
\newcommand{\kms}{\protect \hbox {km s$^{-1}$}}
\newcommand{\mjup}{\ensuremath{M_{\mathrm{Jup}}}}           
\newcommand{\msun}{\ensuremath{M_{\odot}}}                   
\newcommand{\lsun}{\ensuremath{L_{\odot}}}                   
\newcommand{\lbol}{\ensuremath{L_{\rm bol}}}                 
\newcommand{\loglbol}{\protect \hbox{log ($\lbol/\lsun$)}}       
\newcommand{\teff}{\ensuremath{T_{\rm eff}}}
\newcommand{\vtan}{\ensuremath{v_{\rm tan}}}
\newcommand{\dphot}{\ensuremath{d_{\rm phot}}}

\newcommand{\jkmko}{\ensuremath{(J-K)_{\rm MKO}}}
\newcommand{\fldg}{\mbox{\textsc{fld-g}}}
\newcommand{\vlg}{\mbox{\textsc{vl-g}}}
\newcommand{\dmag}{$\Delta$mag}
\newcommand{\rchi}{\ensuremath{\tilde\chi^2}}

\newcommand{\short}{\hbox{2MASS J1119$-$1137}}
\newcommand{\shorta}{\hbox{2MASS J1119$-$1137A}}
\newcommand{\shortb}{\hbox{2MASS J1119$-$1137B}}
\newcommand{\shortab}{\hbox{2MASS J1119$-$1137AB}}

\shorttitle{2MASS J1119$-$1137 is a Binary}
\shortauthors{Best, W. M. J. et al}

\slugcomment{\large Accepted by {\it ApJ Letters}, 2017 June 01}

\begin{document}

\title{The Young L Dwarf 2MASS J11193254$-$1137466 is a Planetary-Mass Binary}
\author{William M. J. Best\altaffilmark{1}, 
  Michael C. Liu\altaffilmark{1},
  Trent J. Dupuy\altaffilmark{2},
  Eugene A. Magnier\altaffilmark{1}
}
\altaffiltext{1}{Institute for Astronomy, University of Hawaii, 2680 Woodlawn Drive, Honolulu, HI 96822, USA; wbest@ifa.hawaii.edu}
\altaffiltext{2}{University of Texas at Austin, Department of Astronomy, 2515 Speedway C1400, Austin, TX 78712, USA}

\begin{abstract}

  We have discovered that the extremely red, low-gravity L7~dwarf
  2MASS~J11193254$-$1137466 is a $0.14''$ (3.6~AU) binary using Keck laser guide
  star adaptive optics imaging.  2MASS~J11193254$-$1137466 has previously been
  identified as a likely member of the TW Hydrae Association (TWA).  Using our
  updated photometric distance and proper motion, a kinematic analysis based on
  the BANYAN II model gives an 82\% probability of TWA membership.  At TWA's
  $10\pm3$~Myr age and using hot-start evolutionary models,
  2MASS~J11193254$-$1137466AB is a pair of $3.7^{+1.2}_{-0.9}$~\mjup\ brown
  dwarfs, making it the lowest-mass binary discovered to date.  We estimate an
  orbital period of $90^{+80}_{-50}$ years.  One component is marginally
  brighter in $K$ band but fainter in $J$ band, making this a probable
  flux-reversal binary, the first discovered with such a young age.  We also
  imaged the spectrally similar TWA L7~dwarf WISEA~J114724.10$-$204021.3 with
  Keck and found no sign of binarity.  Our evolutionary model-derived \teff\
  estimate for WISEA~J114724.10$-$204021.3 is $\approx$230~K higher than for
  2MASS~J11193254$-$1137466AB, at odds with their spectral similarity.  This
  discrepancy suggests that WISEA~J114724.10$-$204021.3 may actually be a tight
  binary with masses and temperatures very similar to
  2MASS~J11193254$-$1137466AB, or further supporting the idea that near-infrared
  spectra of young ultracool dwarfs are shaped by factors other than temperature
  and gravity.  2MASS~J11193254$-$1137466AB will be an essential benchmark for
  testing evolutionary and atmospheric models in the young planetary-mass
  regime.

\end{abstract}


\section{Introduction}
\label{intro}
Brown dwarfs with masses $\lesssim$15~\mjup\ and ages $\lesssim$100~Myr lie at a
nexus of astronomical interest.  They represent the lowest-mass and youngest
products of star formation, and as such offer rare empirical tests for
evolutionary and atmospheric models.  They are also the best field analogs to
directly-imaged giant exoplanets, which are far more difficult to directly
observe due to the glare of their host stars.

Brown dwarfs cool continuously as they age, and the resulting
mass-age-luminosity degeneracy makes their physical properties challenging to
infer without constraints on at least two of those three parameters.  The
atmospheres of young brown dwarfs exhibit clear spectral signatures of low
gravity \citep{Cruz:2009gs,Allers:2013hk}, but the age calibration for these
signatures lacks precision better than $\approx$100~Myr
\citep[e.g.,][]{Liu:2016co}.  Some of the lowest-mass objects have been
identified as members of nearby young moving groups
\citep[e.g.,][]{Gagne:2014gp}, which provide much tighter age constraints and
thus more precise mass estimates from evolutionary models than for ordinary
field objects.  Young binaries with small separations are even more useful as
benchmarks, as their orbits can yield model-independent dynamical masses,
providing exacting tests for models \citep[e.g.,][]{Dupuy:2009jq,Dupuy:2017vz}.

2MASS J11193254$-$1137466 (a.k.a.~TWA 42; hereinafter \short) was discovered by
\citet[hereinafter K15]{Kellogg:2015cf} in a search for L and T~dwarfs with
unusual photometry.  \short\ is an L7 dwarf with extremely red optical and
near-IR colors along with spectral signatures of low gravity indicating youth
\citep[K15;][hereinafter K16]{Kellogg:2016fo}.  K16 identified \short\ as a
candidate member of the TW Hydrae Association
\citep[TWA;][]{Webb:1999kf}, whose age implies a mass of only 4.3--7.6~\mjup\
for this object.  \short\ would be one of the two lowest-mass isolated members
of TWA, comparable only to the L7~dwarf WISEA~J114724.10$-$204021.3
\citep[hereinafter WISEA~J1147$-$2040;
$6.6\pm1.9$~\mjup;][]{Schneider:2016iq,Faherty:2016fx}, and among the
lowest-mass free-floating brown dwarfs known.

We are conducting a high angular-resolution imaging survey of nearby brown
dwarfs to identify binaries.  In this Letter we show that \short\ is a nearly
equal-flux binary with component masses in the planetary regime.

\section{Observations}
\label{obs}
We observed \short\ on 2016 November~25~UT using the laser guide star adaptive
optics (LGS AO) system at the Keck~II Telescope
\citep{vanDam:2006hw,Wizinowich:2006hk}.  We used the facility infrared camera
NIRC2 in its narrow field-of-view configuration, using the $R=13.5$~mag field
star USNO-B1.0~0783-0249513 \citep{Monet:2003bw} located 35\arcsec\ from \short\
for tip-tilt correction.  Skies were mostly clear, with $K$-band seeing of
$1.6''$ measured contemporaneously at UKIRT.  We obtained 3 dithered images at
$K$~band in which \short\ appeared to be an equal-flux binary.  We observed
\short\ again on 2017 March 18 UT in $YJHK$~bands using the same configuration,
under clear skies with seeing~$\approx0.5''$ as measured by the differential
image motion monitor (DIMM) at the Canada-France-Hawaii Telescope (CFHT).
Details are in Table~\ref{tbl.keck}.

In addition, we observed WISEA~J1147$-$2040 on 2016 May 18 UT with a similar
configuration using the $R=17.4$~mag field star USNO-B1.0~0693-0264226
\citep{Monet:2003bw} located 66\arcsec\ away for tip-tilt correction, under
clear skies with seeing~$\approx0.5''$ from DIMM.  The target appeared to be a
single object at $0.11''$~resolution.

\begin{figure*}
\begin{center}
  \begin{minipage}{0.242\textwidth}
    \includegraphics[width=1.00\columnwidth]{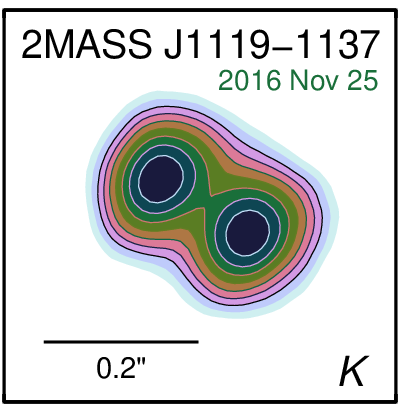}\vspace{12pt}
    \includegraphics[width=0.88\columnwidth]{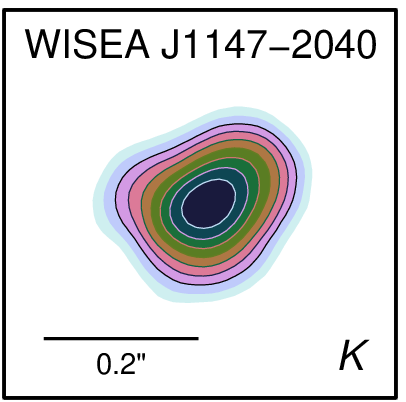}
  \end{minipage}\hspace{-0.22 cm}%
  \begin{minipage}{0.484\textwidth}
    \includegraphics[width=1.00\columnwidth]{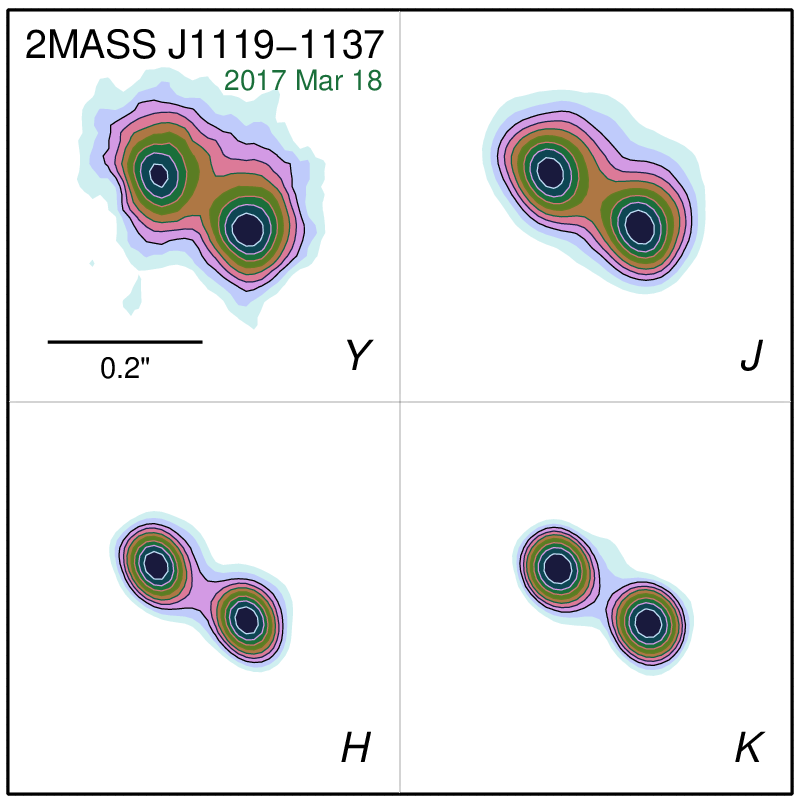}
  \end{minipage}
  \caption{Our Keck LGS AO images of \short\ ({\it top left and 2$\times$2 grid
      at right}) and WISEA~J1147$-$2040 ({\it bottom left}).  North is up and
    East to the left, with filters indicated at lower right and contours marking
    logarithmic intervals from 100\% to 10\% of the peak.  The northeast
    component (\shorta) is marginally brighter in $K$ while the southwest
    component is brighter in $Y$ and $J$, making this system a possible
    flux-reversal binary.}
  \label{fig.images}
\end{center}
\end{figure*}

We reduced and analyzed our data using the methods described in, e.g.,
\citet{Liu:2008ib} and \citet{Dupuy:2017vz}.  Briefly, we calibrated our images
using flat fields and dark frames, performed sky subtraction, and registered and
stacked images to form a final mosaic for each epoch and filter
(Figure~\ref{fig.images}).  We measured the relative astrometry and flux ratios
of the binary by fitting a PSF model to the individual images, applying the
NIRC2 pixel scale, orientation, and distortion correction from
\citet{Service:2016gk}.  For images in which the components were well separated
(all but $Y$ band), we used the StarFinder software package
\citep{Diolaiti:2000dt} to simultaneously solve for an empirical PSF and binary
parameters.  For $Y$ band, we used an analytical PSF of two elliptical
three-component Gaussians.  We applied additional corrections for differential
aberration and atmospheric refraction.  We used the rms of the measurements from
individual images as the uncertainties on the separation, position angle (PA),
and \dmag\ of the binary components (Table~\ref{tbl.keck}), adding the errors in
plate scale (0.4\%) and orientation (0.020~deg) from \citet{Service:2016gk} in
quadrature.

\begin{deluxetable*}{lcccccccccc}
\tablecaption{Keck LGS AO Observations \label{tbl.keck}}
\tabletypesize{\tiny}
\tablewidth{0pt}
\tablehead{
  \colhead{Object} &
  \colhead{Date} &
  \colhead{Filter} &
  \colhead{$N$} &
  \colhead{$t_{\rm int}$} &
  \colhead{Airmass} &
  \colhead{FWHM\tablenotemark{a}} &
  \colhead{Strehl ratio\tablenotemark{a}} &
  \colhead{Separation} &
  \colhead{Position angle} &
  \colhead{\dmag} \\
  \colhead{} &
  \colhead{(UT)} &
  \colhead{} &
  \colhead{} &
  \colhead{(sec)} &
  \colhead{} &
  \colhead{(mas)} &
  \colhead{} &
  \colhead{(mas)} &
  \colhead{(deg)} &
  \colhead{}
}
\startdata
2MASS J11193254$-$1137466   & 2016 Nov 25 & \kmko & 3 & 60   & 1.26 & $94\pm3$       & $0.148\pm0.015$ & $137.8\pm1.7$     & $239.2\pm0.5$     & $0.125\pm0.010$ \\   
                                                    & 2017 Mar 18 & \kmko & 6 & 60   & 1.17 & $59\pm2$       & $0.41\pm0.04$     & $138.10\pm0.10$ & $239.08\pm0.07$ & $0.027\pm0.010$ \\
                                                    &                      & \hmko & 6 & 60   & 1.17 & $51.3\pm0.5$ & $0.212\pm0.011$ & $138.08\pm0.23$ & $238.95\pm0.07$ & $0.008\pm0.007$ \\
                                                    &                      & \jmko  & 6 & 120 & 1.17 & $62\pm9$       & $0.062\pm0.010$ & $137.54\pm0.27$ & $238.72\pm0.17$ & $-0.097\pm0.004$\phs \\
                                                    &                      & \ynirc  & 5 & 120 & 1.18 & $70\pm27$     & $0.036\pm0.008$ & $136.7\pm3.5$     & $238.4\pm0.7$     & $-0.094\pm0.063$\phs \\
WISEA~J114724.10$-$204021.3 & 2016 May 03 & \kmko & 4 & 60   & 1.33 & $108\pm11$   & $0.088\pm0.025$ & \nodata & \nodata & \nodata \\   
\enddata
\tablenotetext{a}{FWHM and Strehl ratios were calculated from each image's
  fitted PSF, except for the $Y$-band data for which we isolated the brighter
  object by rotating the image 180$^\circ$ about the fainter object and
  subtracting.  The tabulated uncertainties are the rms of measurements from
  individual images.}
\end{deluxetable*}

The NIRC2 $J$, $H$, and $K$ filters we used are from the Mauna Kea Observatories
(MKO) photometric system \citep{Simons:2002hh,Tokunaga:2002ex}, and the NIRC2
$Y$-band filter is described in \citet{Liu:2012cy}.  The unresolved photometry
reported in K16 is from the VISTA Hemisphere Survey (VHS; PI McMahon, Cambridge,
UK), which uses MKO $J$ and $H$ filters but a non-MKO $K_{\rm S}$ filter.  We
used the IRTF/SpeX spectrum from K15 to calculate a synthetic
$\kmko=14.658\pm0.066$~mag for \short, flux-calibrated with K16's
$K_{\rm S}$~magnitude.  The VISTA and NIRC2 $Y$ filters are similar enough that
no conversion was necessary.

We split the unresolved \short\ $YJH$ photometry from K16 and our synthetic $K$
magnitude into resolved photometry using our measured flux ratios
(Table~\ref{tbl.prop}).  $K$-band flux decreases monotonically with spectral
type \citep[e.g.,][]{Dupuy:2012bp}, and the northeast component of \short\ is
slightly brighter in $K$, so we designate this object as the ``A'' component.
We note a $0.098\pm0.014$~mag difference between the $K$-band flux ratios
measured at the two epochs.  This may indicate variability in one or both
components, or systematic errors unaccounted in our uncertainties.  We use the
2017 March $K$-band flux ratio for our analysis, as photometry in all other
bands was measured that night, and because of the better image quality.

We also identified a faint source in our $JHK$ images from March 2017, lying
$3.79\pm0.02''$ from \shorta\ at ${\rm PA}=76.47\pm0.13$~deg.  It was not well
detected but appears to be a point source.  Images from the DSS, 2MASS, SDSS,
AllWISE, and Pan-STARRS1 surveys indicate no object at this location.  We
measure flux ratios of 4.8~mag in $J$, 5.0~mag in $H$, and 5.7~mag in $K$,
relative to \shorta.  At the same distance as \short, this source's $J=22.9$~mag
would be consistent with known Y0 dwarfs, but its $\jkmko=1.8$~mag color is
$\gtrsim$3~mag too red \citep{Leggett:2017vj}.  It is almost certainly a
background object.  Its \jkmko\ color suggests an L~dwarf, evolved star, or
galaxy.

\section{Results}
\floattable
\begin{deluxetable}{lccc}
\tablecaption{Properties of 2MASS~J1119$-$1137AB \label{tbl.prop}}
\tablecolumns{4}
\tabletypesize{\scriptsize}
\tablewidth{0pt}
\tablehead{
\colhead{Property} &
\colhead{A component} &
\colhead{B component} &
\colhead{Ref.} \\
\colhead{} &
\colhead{(northeast)} &
\colhead{(southwest)} &
\colhead{}
}
\startdata
\cutinhead{Observed}
R.A.\tablenotemark{a} (deg)   & \multicolumn{2}{c}{169.88521}          & 1 \\
Decl.\tablenotemark{a} (deg) & \multicolumn{2}{c}{$-11.62990$}       & 1 \\
\mua\ (\my)                          & \multicolumn{2}{c}{$-154.0\pm4.0$} & 1 \\
\mud\ (\my)                          & \multicolumn{2}{c}{$-107.9\pm1.8$} & 1 \\
Radial velocity (\kms)            & \multicolumn{2}{c}{$8.5\pm3.3$}       & 2 \\
SpT                                        & \multicolumn{2}{c}{L7 \vlg}                  & 3,4,5 \\
$Y$ (mag)                            & \multicolumn{2}{c}{$19.045\pm0.093$} & 2 \\
$J$ (mag)                             & \multicolumn{2}{c}{$17.330\pm0.029$} & 2 \\
$H$ (mag)                            & \multicolumn{2}{c}{$15.884\pm0.017$} & 2 \\
$K$ (mag)                            & \multicolumn{2}{c}{$14.658\pm0.066$\tablenotemark{b}} & 3 \\
$Y$ (mag)                            & $19.84\pm0.10$ & $19.75\pm0.10$ & 3 \\
$J$ (mag)                             & $18.13\pm0.03$ & $18.04\pm0.03$ & 3 \\
$H$ (mag)                            & $16.59\pm0.02$ & $16.60\pm0.02$ & 3 \\
$K$ (mag)                            & $15.40\pm0.07$ & $15.43\pm0.07$ & 3 \\
$Y-J$ (mag)                         & $1.71\pm0.10$   & $1.71\pm0.10$      & 3 \\
$J-H$ (mag)                         & $1.54\pm0.03$   & $1.43\pm0.03$      & 3 \\
$J-K$ (mag)                         & $2.73\pm0.07$   & $2.61\pm0.07$      & 3 \\
$\Delta(J-K)$ (mag)             & \multicolumn{2}{c}{$0.125\pm0.011$}  & 3 \\
\cutinhead{Estimated}
\dphot\ (pc)                         & \multicolumn{2}{c}{$26.4\pm6.9$}                   & 3,8 \\
$m-M$ (mag)                       & \multicolumn{2}{c}{$2.11\pm0.56$}                 & 3 \\
\vtan\ (\kms)                       & \multicolumn{2}{c}{$23.6\pm6.1$}                   & 3 \\
Projected separation (AU)     & \multicolumn{2}{c}{$3.6\pm0.9$}                     & 3 \\
Semi-major axis (AU)           & \multicolumn{2}{c}{$3.9^{+1.9}_{-1.4}$}               & 3 \\
Orbital Period (yr)                 & \multicolumn{2}{c}{$90^{+80}_{-50}$}                  & 3 \\
\loglbol\ (unresolved)         & \multicolumn{2}{c}{$-4.44^{+0.21}_{-0.27}$}         & 3,6\\
\loglbol\ (resolved)             & $-4.73^{+0.21}_{-0.27}$ & $-4.74^{+0.21}_{-0.27}$  & 3 \\
\cutinhead{Model-derived (Lyon/DUSTY) assuming TWA membership}
Age (Myr)                             & \multicolumn{2}{c}{$10\pm3$~Myr}                  & 7 \\
Mass (\mjup)                       & $3.7^{+1.2}_{-0.9}$        & $3.7^{+1.2}_{-0.9}$         & 3 \\
\teff\ (K)                              & $1013^{+122}_{-109}$   & $1006^{+122}_{-109}$     & 3 \\
\cutinhead{Model-derived (Lyon/DUSTY) assuming young field (\vlg)}
Age (Myr)                             & \multicolumn{2}{c}{$10-100$~Myr}                   & 8 \\
Mass (\mjup)                       & $9.2^{+2.3}_{-1.9}$        & $9.0^{+2.4}_{-1.9}$         & 3 \\
\teff\ (K)                              & $1065^{+133}_{-118}$   & $1059^{+133}_{-118}$     & 3 \\
\enddata
\tablecomments{$JHK$ photometry is on the MKO system.  $Y$ photometry is from
  similar filters on VISTA (integrated light) and Keck/NIRC2 (resolved); no
  conversion was performed.}
\tablenotetext{a}{Epoch 54858.45 (MJD).}
\tablenotetext{b}{Synthetic photometry based on the SpeX prism spectrum
  (K15) and VHS $K_{\rm S}$ photometry (K16).}
\tablerefs{(1) \citet{Best:2017tj}, (2) K16, (3) this work, (4)
  K15, (5) \citet{Gagne:2017gy}, (6) \citet{Faherty:2016fx},
  (7) \citet{Bell:2015gw}, (8) \citet{Liu:2016co}.}
\end{deluxetable}

\subsection{\shortab\ is comoving}
\label{comoving}
\begin{figure}
\begin{center}
    \includegraphics[width=1.0\columnwidth]{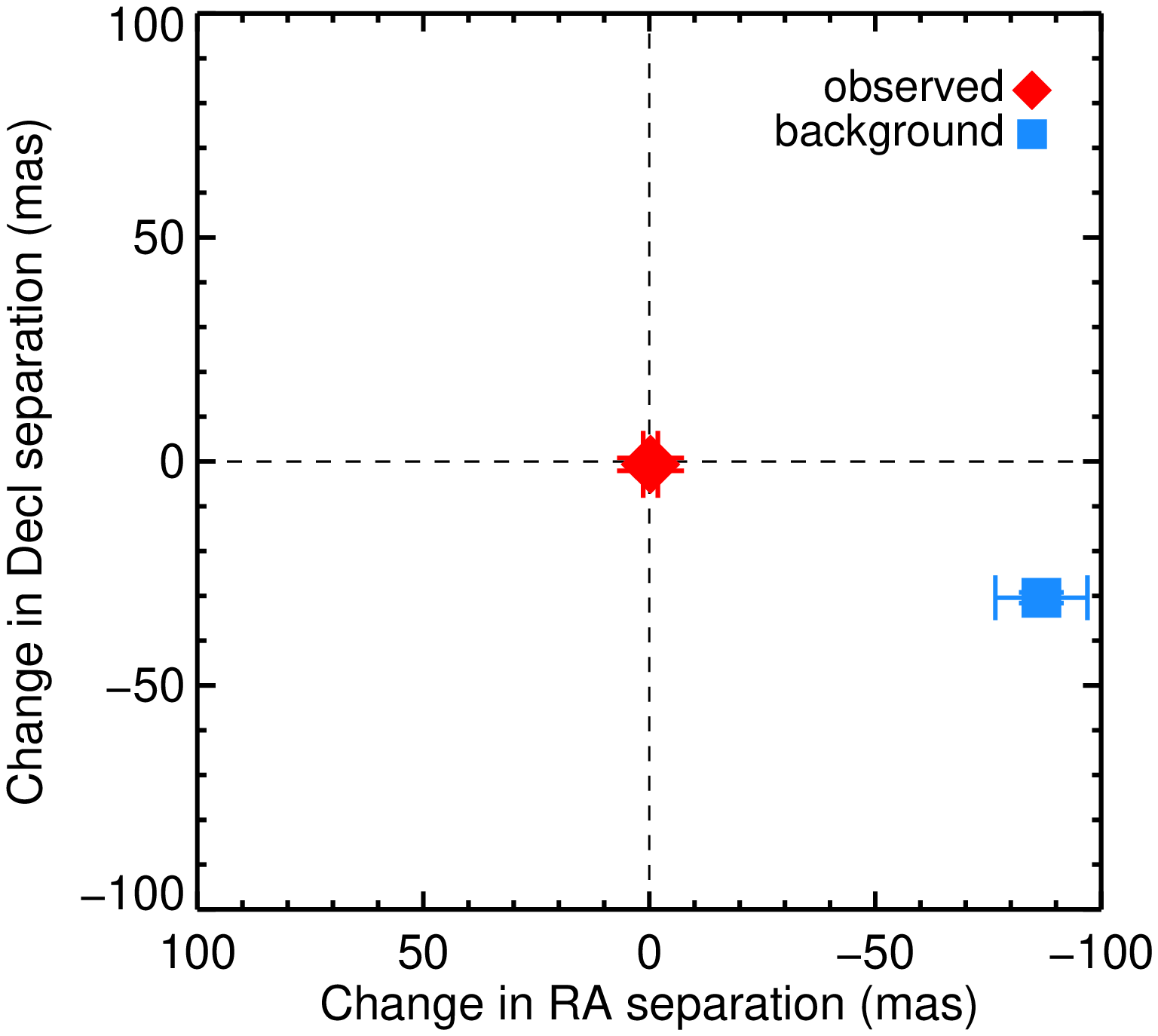}
    \caption{The change in position of \shortb\ with respect to \shorta\ between
      2016 November 25 and 2017 March 18 (red diamond), compared with the change
      expected due to parallax and proper motion of \shorta\ if \shortb\ were a
      stationary background object (blue square).  The error for the background
      position is dominated by the uncertainty on our photometric distance.  The
      observed positions of \shortb\ are consistent with no change, while the
      background position differs by $9.4\sigma$, confirming that \shortab\ is a
      binary.}
  \label{fig.comoving}
\end{center}
\end{figure}

Our $JHK$-band astrometry from March 2017 has a mean separation of
$137.88\pm0.34$~mas and PA~$238.91\pm0.20$~deg, with uncertainties estimated as
in Section~\ref{obs}.  The change in separation from November 2016 is
$0.7\pm1.5$~mas, consistent with no change.  Using the proper motion and
photometric distance of \short\ (Section~\ref{distance}), if \shortb\ were a
stationary background object the separation would have decreased by
$91.7\pm9.8$~mas in March 2017, inconsistent by $9.4\sigma$ from our observation
(Figure~\ref{fig.comoving}).  In addition, images from DSS, 2MASS, SDSS, and
Pan-STARRS1 showed no objects that could appear as a false close companion given
the proper motion.  We therefore conclude that \shortab\ is a gravitationally
bound binary.

\subsection{Spectral type and gravity classification}
\label{spt}
\begin{figure*}
\begin{center}
  \includegraphics[width=2\columnwidth]{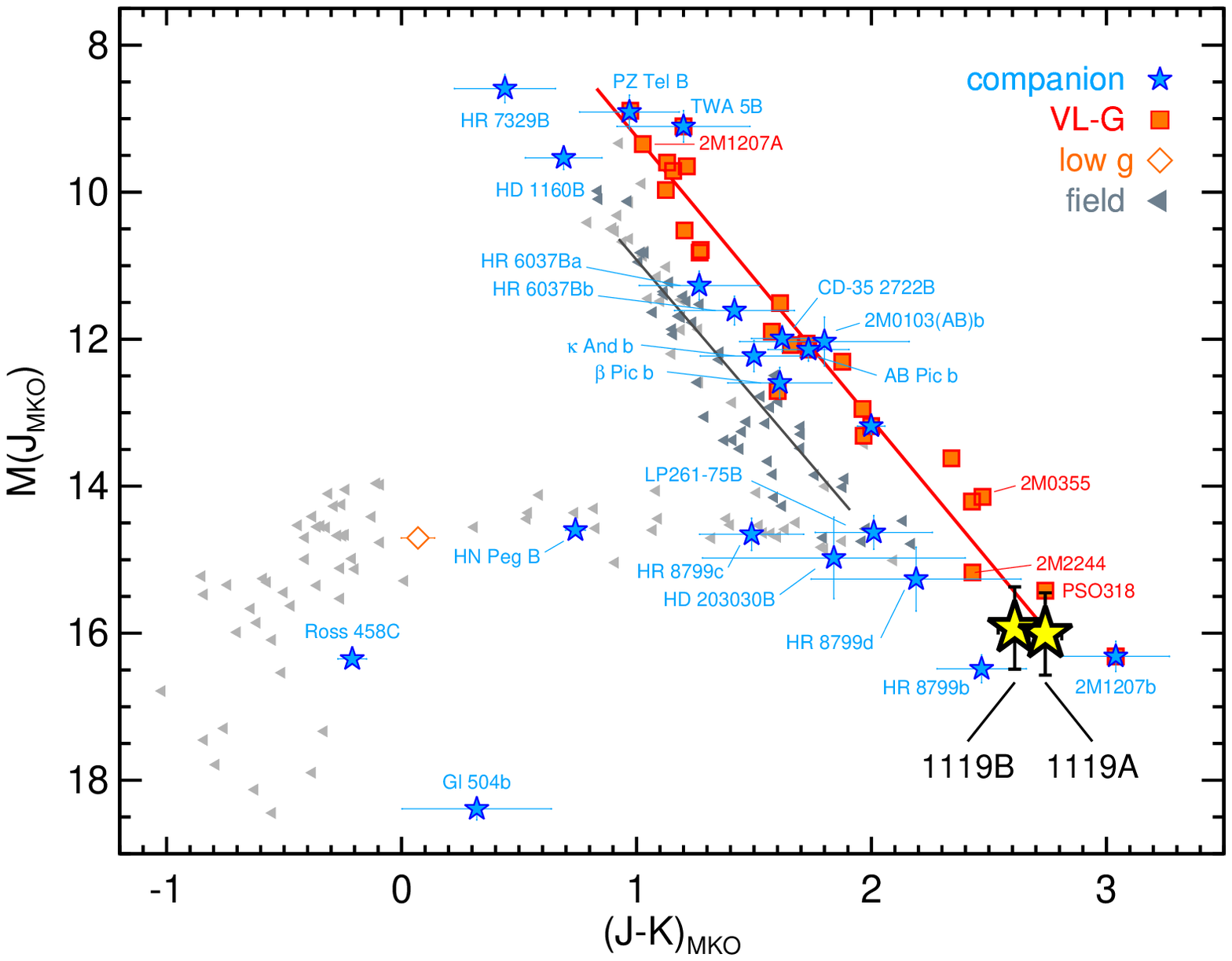}
  \caption{\jmko\ vs. \jkmko\ color-magnitude diagram for ultracool dwarfs
    having parallaxes \citep[adapted from][]{Liu:2016co}.  Gray triangles
    indicate field brown dwarfs, red squares indicate late-M and L dwarfs with
    \vlg\ classifications, and blue stars indicate companions.  Gray and red
    lines show linear fits for the \fldg\ and \vlg\ dwarfs, respectively.
    \shortab\ (yellow stars, using our photometric distance) lies among the
    faintest and reddest planetary-mass L dwarfs.  The $M_{J_{\rm MKO}}$ errors
    for \shortab\ are dominated by the $d_{\rm phot}$; we measure
    $\Delta\jmko=-0.097\pm0.004$~mag.  The relative positions of \shortab\ imply
    that \shortb\ has begun the transition from a red L~dwarf to a bluer
    T~dwarf.}
  \label{fig.cmd}
\end{center}
\end{figure*}

We used the IRTF/SpeX prism spectrum for \short\ and the method of
\citet{Allers:2013hk} to determine a spectral type of L7, concurring with
previous work \citep[K15,][]{Faherty:2016fx,Gagne:2017gy}.  A higher-resolution
$J$-band spectrum (K16) shows weakened K~I absorption lines having equivalent
widths consistent with \vlg\ classification for L7~dwarfs
\citep{Allers:2013hk,Gagne:2017gy}.  We therefore adopt L7~\vlg\ as the
unresolved spectral type.  The extremely red \jkmko\ colors of \shortab\ are
consistent with other low-gravity late-L~dwarfs (Figure~\ref{fig.cmd}).  Both
the \jkmko\ colors and \kmko\ magnitudes for the two components are similar, and
the integrated-light spectrum shows no peculiarities that would suggest a blend
of two objects with different spectral types.  We conclude that both \shorta\
and \shortb\ have spectral types L7~\vlg.

\subsection{Distance}
\label{distance}
We used the spectral-type-to-$M_{K_{\rm MKO}}$ relation for \vlg\ dwarfs from
\citet{Liu:2016co} and our resolved photometry to estimate photometric distances
of $26.3\pm6.8$~pc and $26.6\pm6.9$~pc for \shorta\ and \shortb, respectively.
We adopt $26.4\pm6.9$~pc as the distance to the system, giving the binary a
projected separation of $3.6\pm0.9$~AU.  Our distance is consistent with K16's
kinematic distance of $28.9\pm3.6$~pc assuming membership in TWA.

\subsection{TWA Membership}
\label{twa}
K16 identified \short\ as a candidate TWA member, using the BANYAN~II online
tool \citep{Malo:2013gn,Gagne:2014gp} to calculate an 88\% membership
probability with a contamination probability of 0.003\% based on \short's sky
position, proper motion, radial velocity, and youth.  \citet{Faherty:2016fx}
analyzed moving group membership using four different tools and found $>$90\%
probabilities for TWA in three cases; LACEwING \citep{Riedel:2017dg} found a
contrasting probability of 16\%.  \citet{Gagne:2017gy} used the full BANYAN~II
analysis (including photometry) to calculate a 97\% probability of membership.
\citet{Liu:2016co} found that objects lacking parallaxes which have full
BANYAN~II probabilities $\gtrsim$80\% tend to have memberships confirmed by
subsequent parallax measurements for the well-established moving groups,
including TWA.

We reassessed the moving group membership of \short\ using the BANYAN~II online
tool.  For position and proper motion, we adopted the values from
\citet{Best:2017tj}, which are calculated from 2MASS \citep{Skrutskie:2006hl}
and Pan-STARRS1~$3\pi$ \citep{Chambers:2017vk,Magnier:2017vq} astrometry and
calibrated to the \textit{Gaia}~DR1 reference frame
\citep{GaiaCollaboration:2016cu,Lindegren:2016gr}.
Our \mud\ differs by 35.5~\my\ ($2\sigma$) from K16's value but is consistent
with measurements by K15 and \citet{Gagne:2017gy}.

Using only our astrometry and the radial velocity from K16 as inputs to
BANYAN~II, we obtained a TWA membership probability of 73\%.  The decrease from
K16's 88\% probability is due almost entirely to the difference in \mud.  When
we included our photometric distance, the probability of TWA membership
increased to 82\%.

Based on our BANYAN~II results and the \vlg\ spectrum, \short\ is a very likely
member of TWA.  We note that a comparison in UVW and XYZ of \short\ to known
members of moving groups using \rchi\ as a rubric \citep [see Equation~1
of][]{Liu:2016co} supports membership in TWA, $\beta$~Pictoris (0\% probability
from BANYAN~II), and Tucana-Horologium (0\%), and especially the AB Doradus
(4\%) moving group.  A precise trigonometric distance, now underway by us at
CFHT, is needed to firmly establish the membership.

\subsection{Physical Properties}
\label{results.phys}
As a TWA member, \short\ would share the age of $10\pm3$~Myr derived from the
stellar members \citep{Bell:2015gw}.  We estimated the components' masses and
effective temperatures using bolometric luminosities and the Lyon/DUSTY
hot-start evolutionary models \citep{Chabrier:2000hq}.  To obtain \lbol, we
adjusted the integrated-light value from \citet{Faherty:2016fx} to our new
photometric distance, and decomposed this into individual \lbol\ values using
the binary's \kmko\ flux ratio.  Using these \lbol\ and the assumed age, we then
interpolated a mass and effective temperature from the DUSTY models for each
component.  We propagated the uncertainties on distance, flux ratio, and age
into our calculations via Monte Carlo trials using normal distributions for each
uncertainty, and we quote the resulting median and 68\% confidence limits
(Table~\ref{tbl.prop}).  We estimate masses of $3.7^{+1.2}_{-0.9}$~\mjup\ for
both components.  We did not use the more recent BHAC15 models
\citep{Baraffe:2015fw} as they do not include masses below 0.01~\msun.

If \short\ is not a TWA member, its \vlg\ classification still constrains its
age to $\approx$10--100~Myr \citep{Liu:2016co}.  Our estimated masses and
effective temperatures for this age range (uniformly distributed for error
propagation) are shown in Table~\ref{tbl.prop}.

Using our masses, we considered the effect of \shortab's orbital motion on the
radial velocity measured by K16.  For an edge-on circular orbit with components
at quadrature, we found a difference in radial velocities of
$1.3^{+0.4}_{-0.3}$~\kms\ for TWA masses or $2.0^{+0.6}_{-0.4}$~\kms\ for
field-\vlg\ masses.  The $R\sim6000$ (50~\kms) spectrum of K16 would not resolve
such orbital motion, so the radial velocity from K16 remains valid.

\section{Discussion}
Assuming \short\ is a member of TWA, its total mass is
$7.4^{+2.5}_{-1.9}$~\mjup, making it the lowest-mass binary discovered to date.
The individual masses of \shorta\ and \shortb\ also place them among the
lowest-mass free-floating brown dwarfs, including WISEA~J1147$-$2040 and the Y
dwarfs \citep{Dupuy:2013ks,Leggett:2017vj}.  Even if \shortab\ is actually a
young field object, its total mass of $18.2^{+4.7}_{-3.8}$~\mjup\ would still be
among the lowest-mass binaries, surpassed only by the young binary DENIS-P
J035726.9-441730 \citep[14--15~\mjup;][]{Bouy:2003eg,Gagne:2014gp} and possibly
by the T9+Y0 binaries CFBDSIR~J145829+101343
\citep[$\approx$18--45~\mjup;][]{Liu:2011hb} and WISE~J014656.66+423410.0
\citep{Dupuy:2015dz}.  As an extremely low-mass young binary, \shortab\ will be
a crucial benchmark for tests of evolutionary and atmospheric models.

The isolation of \shortab\ strongly suggests that it is a product of normal
star-formation processes, which therefore must be capable of making binaries
with $\lesssim$5~\mjup\ components.  \shortab\ could be a fragment of a
higher-order system that was ejected via dynamical interactions
\citep{Reipurth:2015dz}, although the lack of any confirmed member of TWA within
10$^\circ$ (projected separation $\approx5$~pc) of \short\ makes this scenario
unlikely.  Formation of very low mass binaries in extended massive disks around
Sun-like stars followed by ejection into the field has been proposed by, e.g.,
\citet{Stamatellos:2009kg}, but disks of this type have not been observed.

Binary brown dwarfs can be monitored to map their orbits, which yield dynamical
masses that stringently test evolutionary models.  We estimated \shortab's
orbital period using Kepler's Third Law.  We first used the projected separation
and a conversion factor from \citet[Table 6, assuming moderate discovery bias
for very low-mass visual binaries]{Dupuy:2011ip} to estimate a semi-major axis
of $3.9^{+1.9}_{-1.4}$~AU.  Our model-derived masses assuming TWA membership
give an orbital period of $90^{+80}_{-50}$~yr.  A dynamical mass can be
determined once $\approx$1/3 of an orbit has been observed
\citep[e.g.,][]{Dupuy:2017vz}, so \shortab\ would yield a dynamical mass in
$\approx$15--55~years.  Assuming a field \vlg\ age, we estimate an orbital
period of $60^{+50}_{-30}$~yr, yielding a mass in $\approx$10--35~years.

The integrated-light spectrum and photometry of \short\ are notably similar to
those of the young L7~dwarf WISEA~J1147$-$2040 \citep{Gagne:2017gy}, implying
similar temperatures and gravity.  However, using the \lbol\ for
WISEA~J1147$-$2040 from \citet{Faherty:2016fx} and the method from
Section~\ref{results.phys}, we estimate a \teff\ of $1242^{+73}_{-69}$~K, which
is $\approx$230~K higher than our estimates for \shorta\ and \shortb.
This discrepancy is particularly surprising given that both objects are very
likely members of TWA and therefore should have the same age and composition.
The simplest resolution is that WISEA~J1147$-$2040 is also an equal-flux binary,
unresolved in our images, with component temperatures very similar to those of
\shortab.  Using the method from Section~\ref{distance} we calculated
$d_{\rm phot}=27.3\pm6.9$~pc for a hypothetical equal-flux binary
WISEA~J1147$-$2040, leading to a maximum projected separation of $2.9\pm0.8$~AU
to remain unresolved in our images.  Another possibility is that \shortab\ is
not a member of TWA and is older; our \vlg-age temperature estimates are only
$\approx$50~K higher, but would then agree with the WISEA~J1147$-$2040 estimate
within uncertainties.  A third intriguing possibility is that the low-resolution
spectra of young red L~dwarfs are driven at least partially by factors other
than temperature and gravity \citep{Allers:2013hk,Liu:2016co}, allowing coeval
objects with differing masses and effective temperatures to have similar
spectra.

Figure~\ref{fig.cmd} compares the \jmko\ vs. \jkmko\ position of \shortab\ to
the ultracool dwarf population, highlighting other low-gravity objects and
substellar companions.  \shortab\ lies among other planetary-mass objects at the
faint red end of the L dwarf sequence.  \shortb\ is brighter in $J$ band and
slightly fainter in $K$ band, making the system a probable flux-reversal binary.
(The $K$ magnitudes for the two components are formally consistent within
uncertainties, but \shortb\ is fainter in all nine of our individual $K$-band
images.)  The flux-reversal phenomenon is a hallmark of field-age L/T transition
binaries \citep[e.g.,][]{Gizis:2003fj,Liu:2006ce,Dupuy:2015gl}, thought to occur
when the cooler component reaches a temperature at which the clouds that
suppress near-IR flux in L~dwarfs begin to clear, reducing the $J$-band opacity
relative to the warmer component \citep[e.g.,][]{Burrows:2006ia}.  The slightly
bluer \jkmko\ color of 2MASS~J1119$-$1137B implies that it is cooler than its
primary.  In field-age brown dwarfs this transition to bluer $J-K$ colors is
typically seen at warmer temperatures \citep[$\approx$1400~K;
e.g.,][]{Dupuy:2012bp}.  The potential flux-reversal of \shortab\ suggests that
it is beginning the transition at $\teff\approx1000\pm100$~K, an even lower
temperature than the $\approx$1100$-1200$~K found for other low (but somewhat
higher) mass L~dwarfs \citep[e.g.][]{Metchev:2006bq,Barman:2011fe,Liu:2013gy},
implying a possible systematic correlation between mass and L/T transition
temperature.  Precise photometry and resolved spectroscopy of \shortab\ with the
Hubble and James Webb Space Telescopes will enable differential studies of the
atmospheres of young planetary-mass objects and may yield insights into the L/T
transition at young ages.

\vspace{20 pt}

We thank the referee for an immediate and helpful review.
The IRTF/SpeX spectrum for \short\ was retrieved from the SpeX Prism Library,
maintained by Adam Burgasser at http://www.browndwarfs.org/spexprism.  The
Pan-STARRS1 Surveys have been made possible through contributions of the
Institute for Astronomy, the University of Hawaii, the Pan-STARRS Project
Office, the Max-Planck Society and its participating institutes, the Max Planck
Institute for Astronomy, Heidelberg and the Max Planck Institute for
Extraterrestrial Physics, Garching, The Johns Hopkins University, Durham
University, the University of Edinburgh, Queen's University Belfast, the
Harvard-Smithsonian Center for Astrophysics, the Las Cumbres Observatory Global
Telescope Network Incorporated, the National Central University of Taiwan, the
Space Telescope Science Institute, the National Aeronautics and Space
Administration under Grant No. NNX08AR22G issued through the Planetary Science
Division of the NASA Science Mission Directorate, the National Science
Foundation under Grant No. AST-1238877, the University of Maryland, Eotvos
Lorand University (ELTE), and the Los Alamos National Laboratory.  This work has
made use of data from the European Space Agency (ESA) mission {\it Gaia}
(\url{http://www.cosmos.esa.int/gaia}), processed by the {\it Gaia} Data
Processing and Analysis Consortium (DPAC,
\url{http://www.cosmos.esa.int/web/gaia/dpac/consortium}).
WMJB and MCL received support from NSF grant AST-1518339.
WMBJ, MCL, and EAM received support from NSF grant AST-1313455.
Finally, the authors acknowledge the significance of  
the summit of Maunakea for
the Native Hawaiian community. We are fortunate to have the opportunity
to conduct observations from this mountain.

\facility{Keck, Pan-STARRS1, {\it Gaia}}

\bibliography{\string~/Astro/LaTeX/willastro}

\end{document}